\documentclass[%
reprint,
showpacs,
floatfix,
amsmath, 
amssymb, 
aps, 
prapplied,
longbibliography
]{revtex4-1}
\usepackage{graphicx}
\usepackage{dcolumn}
\usepackage{bm}
\usepackage{natbib}
\usepackage{hyperref}
\usepackage{float}
\usepackage{mathrsfs}
\usepackage{fancyhdr}
\AtBeginDocument{%
    \newwrite\bibnotes
    \def\bibnotesext{Notes.bib}
    \immediate\openout\bibnotes=\jobname\bibnotesext
    \immediate\write\bibnotes{@CONTROL{REVTEX41Control}}
    \immediate\write\bibnotes{@CONTROL{%
    apsrev41Control,author="08",editor="1",pages="1",title="0",year="1"}}
     \if@filesw
     \immediate\write\@auxout{\string\citation{apsrev41Control}}%
    \fi
}%
\hypersetup{colorlinks=true, citecolor=blue, urlcolor=blue, linkcolor=blue}
\begin{document}
\lhead{}
\chead{}
\rhead{}
\lfoot{}
\cfoot{\thepage}
\rfoot{}

\title{Single-Molecule Doped Crystalline Nanosheets for Delicate Photophysics Studies and Directional Single-Photon Emitting Devices}

\author{Shangming Wei}

\author{Penglong Ren}

\author{Yong He}

\author{Pu Zhang}

\author{Xue-Wen Chen} 
\email{xuewen\_chen@hust.edu.cn}

\affiliation{School of Physics, Huazhong University of Science and Technology, Luoyu Road 1037, Wuhan, 430074, People's Republic of China}%

\date{\today}
\begin{abstract}
Single molecules in solids have been considered as an attractive class of solid-state single quantum systems because they can be chemically synthesized at low cost to have stable narrow transitions at desired wavelengths. Here we report and demonstrate single dibenzoterrylene molecules in crystalline anthracene nanosheets as a robust and versatile solid-state quantum system for delicate photophysics studies and as building blocks of quantum photonic devices. The high-quality nanosheet sample enables robust studies of single-molecule photophysics at room temperature, including for the first time direct observation of single molecule insertion site change, measurement of its associated changes of dipole moment orientation and magnitude, unambiguous determination of excitation-power dependent intersystem crossing rate and triplet lifetime. Thanks to the excellent mechanical stability of the nanosheet sample, we demonstrate its flexible assembly into planar antenna devices to achieve bright Gaussian emission pattern. The thin thickness, good photostability and mechanical stability make the dibenzoterrylene-in-anthracene nanosheet system an excellent candidate as static quantum nodes in integrated photonic circuit.   
\end{abstract}


\maketitle

\section{\label{sec:level1} Introduction}
Solid-state single quantum emitters have become an import resource in quantum information science and technology because they can function as ideal sources of single photons and strong nonlinear elements; owing to the fermionic nature of two-level quantum systems \cite{RN15,RN34}. Since the solid-state platform is compatible with fast-developing nanofabrication techniques, they hold the promise of scaling up the system size together with fine control over the light-emitter interaction at single-photon-single-atom level \cite{RN34,RN90,7479523}. In the past decades, a vast variety of solid-state systems have been investigated, such as quantum dots \cite{RN92,Loredo:16,PhysRevLett.116.020401}, color centers in diamond \cite{PhysRevLett.85.290,RN118,RN86}, defects in two-dimensional materials \cite{doi:10.1002/adma.201606434} and organic molecules in crystalline hosts \cite{RN49}, to name a few. Among them, single molecules embedded in crystalline matrix host, though being less popular in the field of quantum information science, are in fact a class of very versatile system \cite{RN15,RN49,RN76,RN79}, Most of them belong to the family of polycyclic aromatic hydrocarbons (PAH) \cite{RN124} and can be chemically synthesized at low cost to have stable emission at desired wavelengths. In particular, 7.8,15.16-dibenzoterrylene (DBT) molecules embedded in suitable crystalline matrices such as anthracene (AC) have been reported as a very promising systems due to near-unity quantum yield, non-blinking emission, good photostability, and lifetime-limited emission linewidth at cryogenic temperatures \cite{RN22,RN28,RN24,PhysRevX.7.021014,RN23,RN82}. There have been already attempts to integrate DBT molecules in crystalline matrices into planar photonic circuit for on-chip single-photon emission \cite{RN74,RN14,RN119} and coherent interactions \cite{RN70,RN80}. However, the employed fabrication strategies and sample formats are not easily compatible with the requirements of cryogenic-temperature operation or/and controlled interaction with planar photonic elements on the chip. From a different perspective, the properties of a molecule depend not only on its own chemical structure but also on its local environment of the matrix. Single molecules can be used as probes to investigate the structure and dynamics of condensed matter and mechanical vibrations at the nanometer scale \cite{RN76,RN35,Moerner1670}. Thus studies of photophysics of single molecule in high-quality crystal matrix could lay the ground for the applications as an ultrasensitive sensor \cite{RN76}.

\par In this work, we demonstrate the growth of DBT doped crystalline anthracene (DBT:AC) nanosheets for delicate single-molecule photophysics studies and building blocks of single-photon quantum devices. Specifically, we report the first real-time observation of molecule insertion site jump and the associated changes of the emission dipole (both orientation and magnitude) at room temperature. In addition, we deduce the excitation-power dependent intersystem crossing rate and triplet state lifetime of DBT molecules at room temperature by measuring the second-order photon correlation over a large dynamic range (0.1ns $\sim$ 0.1 ms) \cite{RN1,RN8}. The thin and mechanically rigid DBT:AC nanosheet sample ensures good coupling, either via direct or evanescent field coupling, and easy assembly with photonic nanostructures through micro-manipulation techniques. As an example, we integrate the DBT:AC nanosheet into a planar antenna structure \cite{RN13,RN57,RN58} and demonstrate a stable and bright single-photon emitting device with a Gaussian emission pattern.
\\
\\

\section{\label{sec:level1} Experimental results and discussion}
\subsection{\label{sec:level2} Sample growth, characterization and setup}

\par We begin the discussion with the growth of crystalline DBT:AC nanosheets. Figure 1(a) shows the home-built apparatus for sample growth through a co-sublimation process \cite{RN22}. The well-mixed solids of anthracene and DBT (mixing ratio $10^{6}:1$) in a glass tube is heated to $255^{\circ}$C under nitrogen atmosphere and then a clean glass coverslip is placed above the opening of glass tube for a few minutes to let the mixture vapor to condense and form crystalline anthracene nanosheets doped with DBT molecules \cite{supplemental}. The nanosheets are hexagons as displayed by an atomic force microscope (AFM) topography image in Fig. 1(b) and also an optical microscopy image in the inset of Fig. 1(c). The cross section along the dashed line in the AFM image indicates a very flat crystal surface with sharp edges and a typical thickness of 100 nm. The nanosheets are mechanically rigid and robust. By using an optical fiber tip mounted on a three dimensional translation stage, we manage to freely micro-manipulate the nanosheets, for instance, transferring them to other substrates or photonic structures as sketched in Fig. 1(a) in a controlled fashion. Single-molecule spectroscopy experiments are conducted for DBT:AC nanosheets on coverslip with a home-built inverted microscope 
\begin{figure}[h]
\centering
\includegraphics[width=8.6cm]{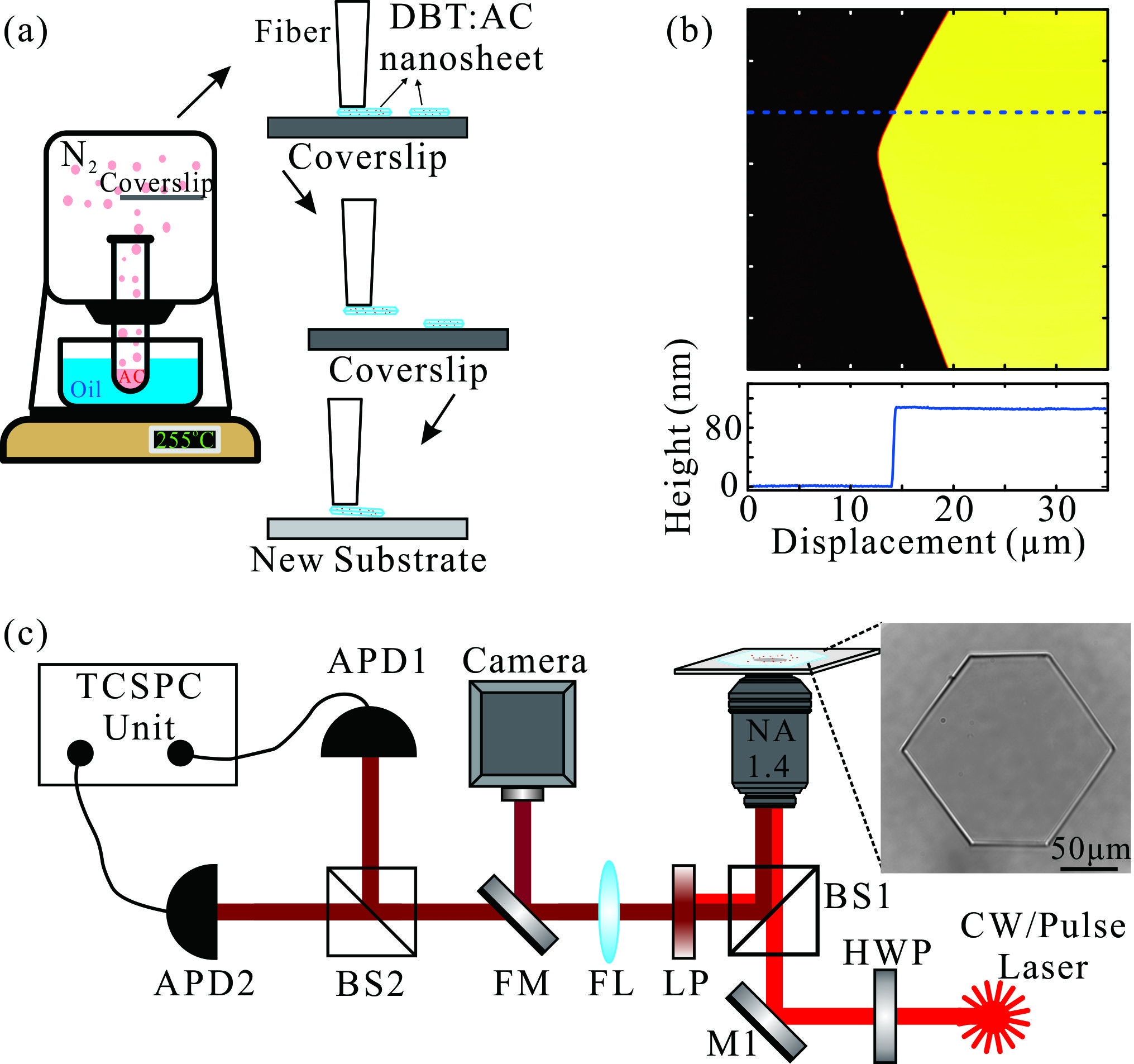}
\caption{Experimental setup and sample characterization. (a) Left: apparatus for DBT:AC nanosheet sample growth, Right: schematics of micro-manipulation of the nanosheets. (b) AFM topography image of part of a nanosheet. Lower panel is the plot of the height along the cross section. (c) Homebuilt inverted microscope setup for single-molecule spectroscopic measurements. LP: longpass filter, FM: flip mirror, FL: flip lens and HWP: halfwave plate.}
\label{fig1}
\end{figure}
\noindent setup as illustrated in Fig. 1(c). The microscope with an oil-immersion objective of NA=1.4, various detectors and time-correlated single-photon counting (TCSPC) unit, provide access to a wide range of optical measurements, including photoluminescence (PL) time trace and decay dynamics, second-order photon correlation function $g^{2}(\tau)$, back focal plane (BFP) imaging and fluorescence saturation. A continuous wave (CW) or pulsed laser at 720 nm ($\sim$30 ps, 2.1 MHz repetition rate) with a diffraction-limited spot is launched into the system to address single DBT molecules \cite{supplemental}.

\begin{table*}
\centering
\caption{\label{tab:table1} Summary of measured molecules with insertion site change from B to A}
\begin{ruledtabular}
\begin{tabular}{ccccc}
\multicolumn{1}{c}{   }&
\multicolumn{1}{c}{Emission peak (nm)}&
\multicolumn{1}{c}{Emission width (nm)}&
\multicolumn{1}{c}{Lifetime (ns)}&
\multicolumn{1}{c}{Saturation intensity (kcps)}\\
Molecule & (Site B / A) & (Site B / A) & (Site B / A) & (Site B / A)\\
\hline
M1 & 789 / 780 & 37 / 21  & 3.99 / 4.93  & 747 / 559 \\
M2 & 787 / 782  & 32 / 23  & 4.07 / 4.96  & 718 / 624 \\
M3 & 786 / 780  & 37 / 22  & 4.05 / 4.92  & 830 / 626 \\
M4 & 787 / 780  & 35 / 20  & 4.08 / 4.97  & 797 / 605 \\
\end{tabular}
\end{ruledtabular}
\end{table*}

\subsection{\label{sec:level2} Light-driven molecule insertion site jump}
\par The high-quality crystalline DBT:AC nanosheets provide an ideal platform for studying delicate single-molecule photophysics, for instance, direct observation of insertion site change at ambient conditions. In host-guest type of single-molecule systems, the guest molecule replaces one or a few adjacent host molecules, which usually forms one or more than one kind of well-defined insertion sites. All past experimental observations and studies of single molecules with different insertion sites, for instance, terrylene in p-terphenyl \cite{RN120} and DBT molecules in anthracene \cite{RN21}, have been conducted under cryogenic temperatures in order to have stable environment and narrow emission with low background. Here, for the first time, we report real-time observation of insertion site change of single DBT molecules in anthracene nanosheet at room temperature and provide quantitative measurements of the associated changes of the emission dipole, including both the orientation and magnitude. 
\begin{figure}
\centering
\includegraphics[width=8.6cm]{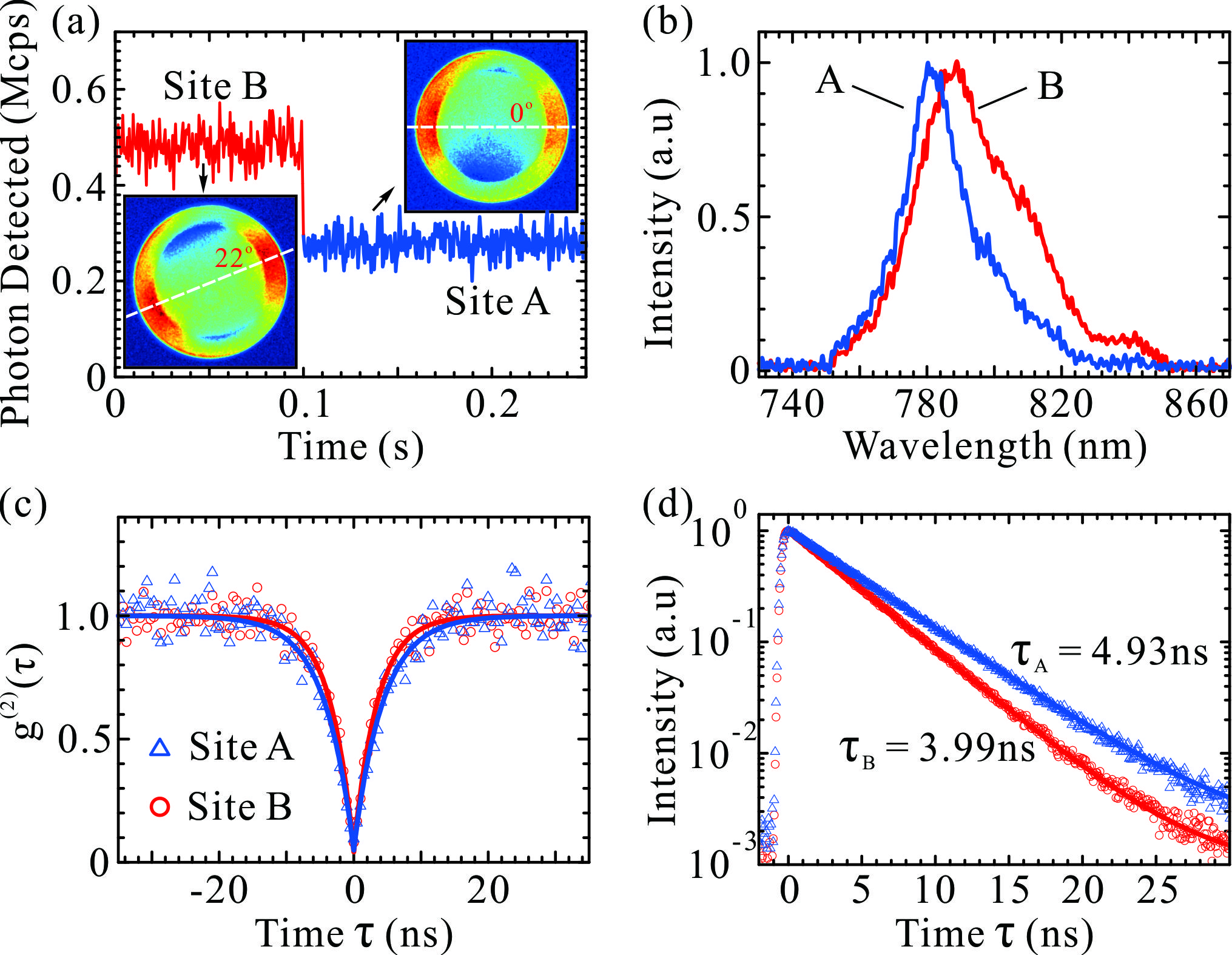}
\caption{Real-time observation of molecule insertion site change at room temperature. (a) PL time trace from a single DBT molecule under constant excitation. Insets show the back-focal plane images of the two corresponding time trace segments. (b) Emission spectra, (c) Normalized second-order photon correlation functions and (d) PL decay dynamics of the same molecule at two sites.}
\label{fig2}
\end{figure}

\par Figure 2(a) shows a fraction of the recorded PL time trace from one DBT molecule under constant laser excitation. The fluorescence is non-blinking but experiences one abrupt change of the intensity. The emission spectra of the two segments as shown in Fig. 2(b) are distinct, with a peak wavelength shift of about 9 nm and the linewidth changes from about 37 nm to 21 nm. We observe that the blue shifted emission with narrower linewidth is more stable and the majority ($\sim$90\%) of the DBT molecules belong to this state. These observations are consistent with previous experiment at low temperature for different DBT molecules at two insertion sites, i.e., main and red insertion sites \cite{RN21}. Thus we believe the observed abrupt fluorescence change here corresponds to the insertion site change (labelled as site A and B here) of the same molecule under light excitation. We remark in passing that the high-quality crystalline AC nanosheets are excellent hosts for DBT molecules, whose emission linewidth (site A) is about 1/3 of the linewidth of DBT:AC samples prepared through the spin-casting method \cite{RN28}. The measured second-order photon correlation functions of the molecule at site A and B are depicted in Fig. 2(c), where the pronounced anti-bunching dips at zero delay confirm both emissions are from the same one molecule. The insertion site change is also associated with the change of the emission dipole moment, including both the orientation and magnitude. The emission pattern at BFP is correlated to the orientation of the emission dipole \cite{RN13,RN57,Lieb:04}. The measured BFP patterns of the same molecule at two sites are displayed as insets of Fig. 2(a) and the images indicate that the emission dipoles mainly lie in the plane of the nanosheet. One clearly observes that along with the site change the orientation of the emission dipole changes by $22^{\circ}$ in the plane, as confirmed by theoretical simulations for an electric dipole embedded in an anisotropic matrix on the glass substrate \cite{RN68} (Section 5 of Supplemental Materials \cite{supplemental}). This is the first direct experimental demonstration of the change of emission dipole orientation upon an insertion site change from the same molecule. The magnitude of the emission dipole moment is proportional to the square root of the spontaneous emission rate \cite{loudon2000quantum}, i.e., the inverse of the PL lifetime, if the quantum efficiency of the emission is close to unity and the local density of states doesn’t change, which are the cases for DBT molecules in anthracene matrix (Section 3 of Supplemental Materials\cite{supplemental}). The PL decay curves of the molecule in site B and A are depicted by the red and blue traces in Fig. 2(d), respectively. By fitting the curves with a mono-exponential decay model, one observes that the PL lifetime changes from 4.0 ns of site B to 4.9 ns of site A, corresponding that the magnitude of the dipole moment reduces by about 10\%. We have measured a series of molecules experiencing site changes under constant laser excitation and the results are summarized in Table I. One sees that the observations from different molecules are consistent and reproducible.

\begin{figure*}
\centering
\includegraphics{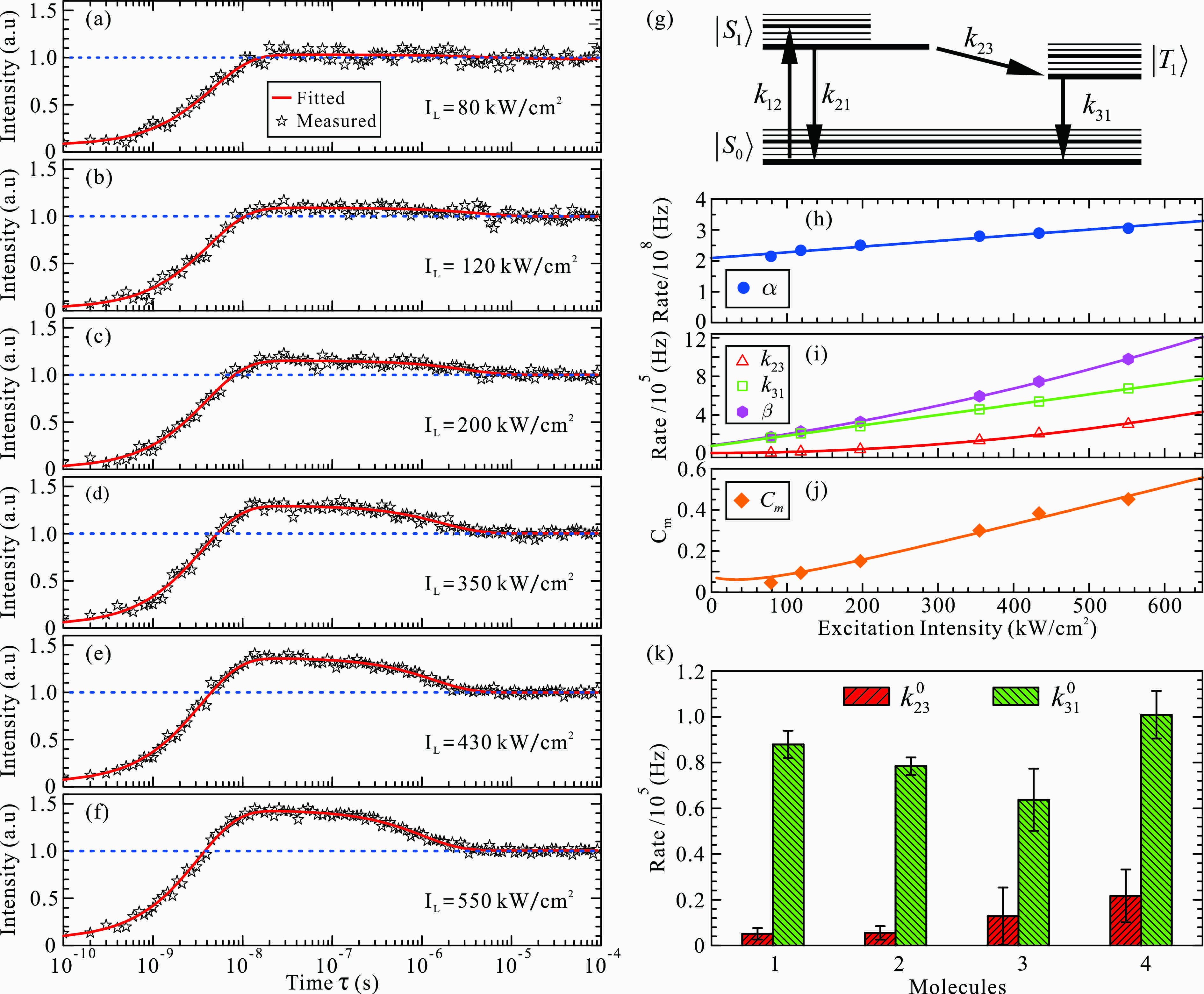}
\caption{Determination of the rates of intersystem crossing and triplet state decay for DBT:AC nanosheet at room temperature. (a)$\sim$(f) normalized second-order photon correlation functions measured at various excitation light intensities. (g) simplified level scheme of the molecule. Deduced $\alpha$(h), $\beta$, $k_{23}$, $k_{31}$ (i) and $C_{m}$ (j) as a function of the excitation light intensity. (k) Summary of $k_{23}$, $k_{31}$ at the limit of zero excitation intensity for four DBT molecules.}
\label{fig3}
\end{figure*}

\subsection{\label{sec:level2} Quantifying intersystem crossing rate and triplet decay}
\par The stable emission of single DBT molecules at site A allows us to make robust measurements on the rates of singlet-triplet intersystem crossing, triplet state decay and their excitation power dependence at room temperature. The existence of triplet state and the intersystem crossing will lead to photon bunching from the single molecule at the delay close to the triplet lifetime \cite{RN1}, which should be carefully characterized. To unambiguously determine these rates, we construct the second-order photon correlation functions $g^{2}(\tau)$(normalized) over a large dynamic range covering from 0.1 nanosecond (ns) to 0.1 millisecond (ms) \cite{supplemental}. The measured $g^{2}(\tau)$ curves are plotted in Fig. 3(a) – 3(f) for a variety of CW excitation light intensities. The correlation function, associated with the photophysics of the molecule, strongly depends on the excitation light intensity. In short-time delay ($\tau<$ 20ns), one observes antibunching statistics due to the emission from the singlet state of the molecule and the exponential rise becomes faster as the excitation intensity increases. For delays from several tens of nanoseconds to a few microseconds, one sees clear photon bunching phenomena due to triplet trapping and intersystem crossing. Moreover, both the bunching level and its decay rate increases as the excitation intensity increases.For long-time delay, $g^{2}(\tau)$ approaches unity as expected.

\par For the whole dynamic range, the normalized second-order photon correlation function can be expressed as 
\begin{equation}
\label{equation1}
	g^{2}(\tau)=a(1-e^{-\alpha\tau})-b(1-e^{-\beta\tau})+c
\end{equation}
where the first term describes the short-term antibunching statistics, the second term is for the bunching effect and the third term c is responsible for the background $(a-b+c=1)$, respectively. A simplified energy level structure is shown in Fig. 3(g). The excitation-intensity dependent constants $\alpha$, $\beta$ and the contrast of photon bunching $C_{m}$ are related to the decay rates as follows \cite{RN1},
\begin{equation}
\label{equation2}
	\alpha=k_{21}+\sigma I_{L}
\end{equation}
\begin{equation}
\label{equation3}
	\beta \approx k_{31}+\sigma I_{L}k_{23}/(\sigma I_{L}+k_{23})	
\end{equation}
\begin{equation}
\label{equation4}
	C_{m}=\sigma I_{L}k_{23}/[k_{31}(\sigma I_{L}+k_{23})]
\end{equation}
 \noindent where $\sigma$ is the absorption cross section of the molecule, $k_{ij}$ is the decay rate from simplified level $i$ to level $j$, and $I_{L}$ is the excitation intensity. By fitting the measured $g^{2}(\tau)$ with the model given by Eq. (1), we can obtain the decay constants and the bunching contrast at various excitation powers (See details in Section 4 of Supplemental Materials\cite{supplemental}). The filled solid symbols in Fig. 3(h), 3(i) and 3(j) depict $\alpha$, $\beta$, $C_{m}$ as a function of the excitation intensity, respectively. From Eq.(2)-(4), one also gets the power-dependent rates of $k_{23}$ and $k_{31}$, which increases dramatically with the excitation power as shown by the triangles and squares in Fig. 3(i), respectively. The dependence of $k_{23}$ and $k_{31}$ on the excitation power may result from pumping the lowest triplet state to higher triplet states\cite{RN8,RN121}, which could make fast reverse intersystem crossing and effectively accelerate both $k_{23}$ and $k_{31}$. We have measured the photon statistics of 4 DBT molecules in anthracene nanosheet and the extracted rates of intersystem crossing and triplet decay at very weak excitation are shown in Fig. 3(k). Our measurements show that the intersystem crossing rate $k_{23}$ is in the order of 5 - 20 kHz at very low excitation power and increases quadratically with the increment of the excitation to typically values of hundreds of kHz for DBT molecules in anthracene nanosheet at room temperature. These values are 2 $\sim$ 4 orders of magnitude larger than the value measured at cryogenic temperature\cite{RN22} but still guarantee intersystem crossing yields well below 0.1\%. The decay rate of the triplet state also increases with the excitation intensity from about 80 kHz at zero excitation light to typical values of several hundred kHz, making the triplet state lifetime of a few microsecond. We remark here that the DBT:AC nanosheet sample at room temperature has low intersystem cross yield and very short trapping times of a few microseconds depending on the level of excitation, which manifest it as a good solid-state quantum system. 

% \begin{figure}[b]
% \centering
% \includegraphics[width=8.6cm]{Figs_4.jpg}
% \caption{Assembly of DBT:AC nanosheet into planar antenna structure for directional single-photon emission. (a) Schematic diagram of the device structure. Inset: optical image of the device seen from the top. (b) Saturation of single DBT molecule emission from the device. Inset: normalized second-order photon correlation function. (c) Measured and (d) simulated back-focal plane (BFP) image of the emission from a single molecule in the device. }
% \label{fig4}
% \end{figure}

\subsection{\label{level2} Assembly for directional single-photon emission}
\par Having systematically studied the emission properties of DBT:AC nanosheets, we next demonstrate them as versatile building blocks of functional single-photon emitting devices. The thin and mechanically rigid DBT:AC nanosheets can be readily assembled to photonic nanostructures via the micro-manipulation technique, for instance, to form a planar antenna device for directional single-photon emission \cite{RN13,RN57,RN58}. Here we follow the design concept of planar Yagi-Uda antenna structure \cite{RN58,RN122} to achieve a bright Gaussian emission pattern from a single molecule. In stark contrast to isotropic matrices, the crystalline anthracene nanosheet is biaxial anisotropic \cite{doi:10.1143/JPSJ.17.113}, and the antenna structure may significantly enhance the effect of anisotropy, which influences the emission pattern \cite{RN68}. We have carefully taken the anisotropy effect into account and designed the parameters for the antenna structure. As the device structure schematically shown in Fig. 4(a), the nanosheets of about 150 nm thick are first transferred to a coverslip coated with 20 nm thick Au layer, then covered with 20 nm thick polyvinyl alcohol film and afterwards deposited 100 nm Au film. An optical microscope image in the inset of Fig. 4(a) shows the edges of a nanosheet inside the device. The stable emission from a single DBT molecule at site A is collected from the side of coverslip and measured for photon statistics, saturation and emission pattern. The fluorescence saturation curve of the molecule is plotted in Fig. 4(b) while its second-order photon function with $g^{2}(0) \approx 0.04$ is shown in the inset. The detected photon rate saturates at 1.5 million counts per second (Mcps), which is about 2.5 times of the value for the standard case where the nanosheet is directly placed on the coverslip. The emission pattern of the molecule is clearly Gaussian as indicated by the BFP image shown in Fig. 4(c). Our simulation in Fig. 4(d) shows that the emission pattern of a molecule embedded in the nanosheet with 40 nm distance from the lower crystal surface agrees well with the measured BFP \cite{supplemental}. The superimposed white circles on the images indicate the corresponding NAs and the calculations show that 66\% of the total emission (including the losses due to absorption in metal) can be collected through an objective with a NA of 0.8. We remark in passing that single DBT molecules in the device are extremely stable and can function properly for months under ambient conditions, probably due to the fact that the present DBT:AC nanosheets in the device are very well isolated from the environment. 
\begin{figure}
\centering
\includegraphics[width=8.6cm]{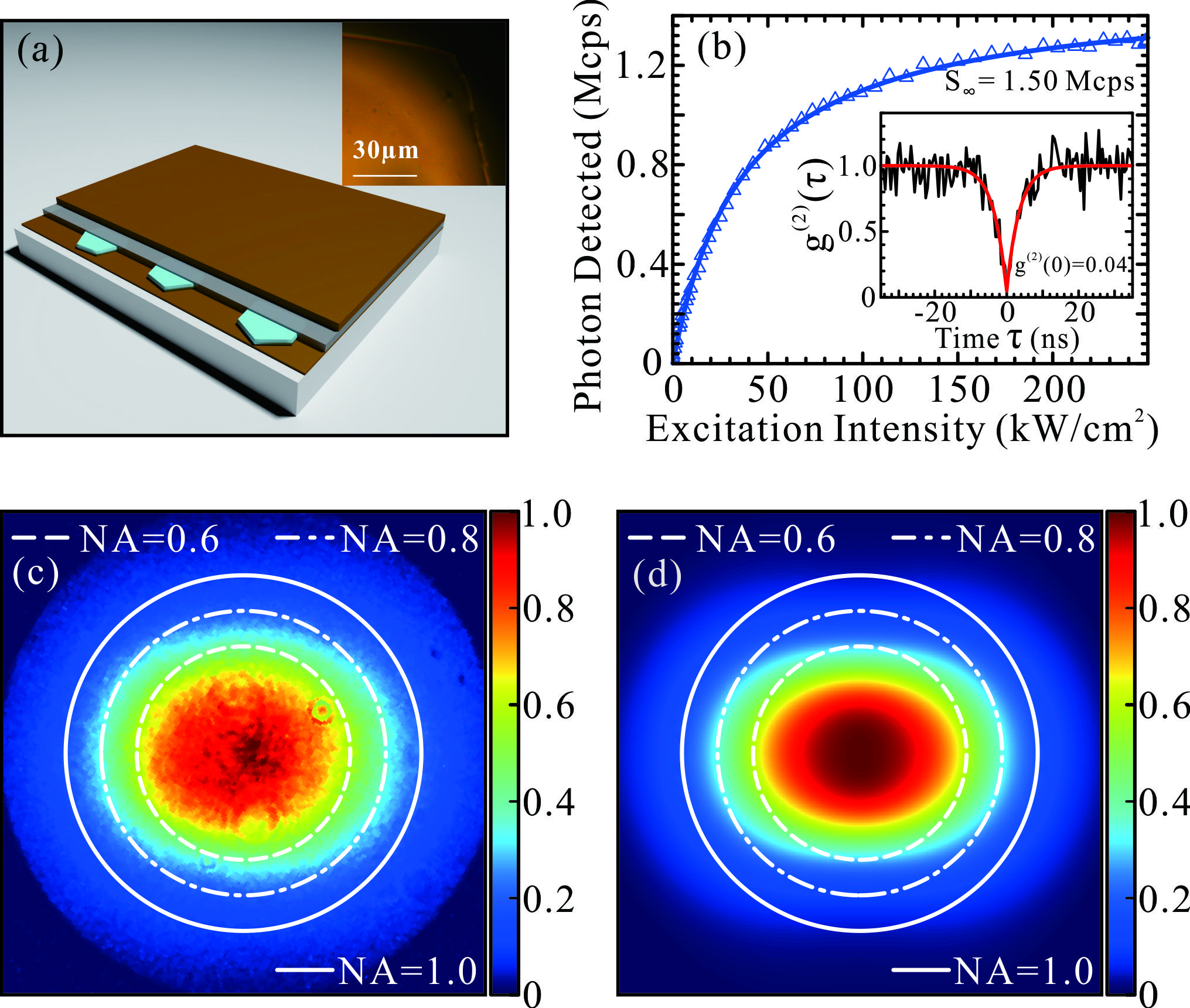}
\caption{Assembly of DBT:AC nanosheet into planar antenna structure for directional single-photon emission. (a) Schematic diagram of the device structure. Inset: optical image of the device seen from the top. (b) Saturation of single DBT molecule emission from the device. Inset: normalized second-order photon correlation function. (c) Measured and (d) simulated back-focal plane (BFP) image of the emission from a single molecule in the device. }
\label{fig4}
\end{figure}

\section{\label{sec:level1} Conclusion and outlook}
\par In summary, we have systematically investigated a class of clean, robust and versatile solid-state single quantum system, i.e., DBT molecules embedded in crystalline anthracene nanosheet. The high-quality DBT:AC nanosheet sample, manifested partially from the narrow emission spectrum, enables delicate single-molecule photophysics studies at room temperature, including real-time observation of insertion site change, the associated changes of emission dipole moment orientation and magnitude, unambiguous determination of excitation-power dependent rates of intersystem crossing and triplet decay of single DBT molecules. Moreover, benefiting from the excellent mechanical stability of the nanosheet sample, we have demonstrated its flexible assembly into planar antenna devices for bright Gaussian emission from single DBT molecules. The thin thickness, the excellent photostability, mechanical stability and cryogenic compatibility make the DBT:AC nanosheet system an excellent candidate as static quantum nodes in integrated quantum photonic circuit \cite{RN34,RN90,7479523}. We expect that versatile forms of coupling of DBT:AC nanosheets with photonic structures opens the door to a number of interesting experiments such as on-chip generation of indistinguishable single photons \cite{RN123}, on-chip Hong-Ou-Mandel interference from independent single emitters \cite{PhysRevLett.104.123605} and quantum photonic nodes for deterministic high-efficiency quantum state transfer \cite{tian2019quantum}.

\begin{acknowledgments}
\par We acknowledge financial support from the National Natural Science Foundation of China (Grant Number 11874166, 11604109, 11474114), the Thousand-Young-Talent Program of China and Huazhong University of Science and Technology.
\end{acknowledgments}
\par S.W. and P.R. contributed equally to this work.

% %%\bibliographystyle{revtex4-1}
% \bibliography{References.bib}
% \end{document}


\end{document}